\documentclass[natbib]{elsart}
\usepackage{amsmath}
\usepackage{amssymb}
\usepackage{amsfonts}
\usepackage{graphicx}
\usepackage{dcolumn}
\usepackage{bm}
\usepackage{color}

\begin{document}

\begin{frontmatter}

\title{Scale invariant distribution and multifractality of volatility multipliers in stock markets}

\author[SB,SS]{Zhi-Qiang Jiang},
\author[SB,SS,RCSE]{Wei-Xing Zhou\thanksref{EM}}
\address[SB]{School of Business, East China University of Science and Technology, Shanghai 200237, China}
\address[SS]{School of Science, East China University of Science and Technology, Shanghai 200237, China}
\address[RCSE]{Research Center of Systems Engineering, East China University of Science and Technology, Shanghai 200237, China}

\thanks[EM]{Corresponding author. {\it E-mail
address:} wxzhou@ecust.edu.cn (W.-X. Zhou)}

\begin{abstract}
The statistical properties of the multipliers of the absolute
returns are investigated using one-minute high-frequency data of
financial time series. The multiplier distribution is found to be
independent of the box size $s$ when $s$ is larger than some
crossover scale, providing direct evidence of the existence of scale
invariance in financial data. The multipliers with base $a=2$ are
well approximated by a normal distribution and the most probable
multiplier scales as a power law with respect to the base $a$. We
unravel that the volatility multipliers possess multifractal nature
which is independent of construction of the multipliers, that is,
the values of $s$ and $a$.
\end{abstract}


\end{frontmatter}

\section{Introduction}
\label{intro} It has been a long history that physicists show
interests on financial markets, which can be at least traced back to
1900 when Bachelier modeled stock prices with Brownian motions
\cite{Bachelier-1900}. In the middle of last century, Mandelbrot
proposed to characterize the tail distributions of income and cotton
price fluctuations with the Pareto-L\'evy law and applied R/S
analysis to investigate the temporal correlations in the evolution
of stock prices \cite{Mandelbrot-1997}. In recent years since the
seminal work of Mantegna and Stanley
\cite{Mantegna-Stanley-1995-Nature}, Econophysics has attracted
extensive interest in the physics community.

As an analogue to turbulence, many time series observed in the
financial markets are reported to possess multifractal properties
\cite{Ghashghaie-Breymann-Peinke-Talkner-Dodge-1996-Nature,Mantegna-Stanley-1996-Nature},
such as the foreign exchange rate
\cite{Ghashghaie-Breymann-Peinke-Talkner-Dodge-1996-Nature,Mantegna-Stanley-1996-Nature,Vandewalle-Ausloos-1998-IJMPC,Schmitt-Schertzer-Lovejoy-1999-ASMDA,Ivanova-Ausloos-1999-EPJB,Baviera-Pasquini-Serva-Vergni-Vulpiani-2001-PA,Muniandy-Lim-Murugan-2001-PA,Xu-Gencay-2003-PA},
gold price \cite{Ivanova-Ausloos-1999-EPJB}, commodity price
\cite{Matia-Ashkenazy-Stanley-2003-EPL}, stock price
\cite{Matia-Ashkenazy-Stanley-2003-EPL,Turiel-Perez-Vicente-2003-PA,Oswiecimka-Kwapien-Drozdz-Rak-2005-APP,Olsen-2000-PP,Turiel-Perez-Vicente-2005-PA,Norouzzadeh-Jafari-2005-PA},
stock market index
\cite{Bershadskii-2001-JPA,Sun-Chen-Wu-Yuan-2001-PA,Sun-Chen-Yuan-Wu-2001-PA,Andreadis-Serletis-2002-CSF,Gorski-Drozdz-Speth-2002-PA,Ausloos-Ivanova-2002-CPC,Balcilar-2003-EMFT,Lee-Lee-2005a-JKPS,Lee-Lee-2005b-JKPS,Lee-Lee-Pikvold-2006-PA,Wei-Huang-2005-PA},
to list a few. Extensive methods have been adopted to extract the
empirical multifractal properties in financial data sets, for
instance, the wavelet transform module maxima (WTMM)
\cite{Holschneider-1988-JSP,Muzy-Bacry-Arneodo-1991-PRL,Muzy-Bacry-Arneodo-1993-PRE}
and the multifractal detrended fluctuation analysis (MF-DFA)
\cite{Kantelhardt-Zschiegner-Bunde-Havlin-Bunde-Stanley-2002-PA}. A
time series of the price fluctuations possessing multifractal nature
usually has either fat tails in the distribution or long-range
temporal correlation or both
\cite{Kantelhardt-Zschiegner-Bunde-Havlin-Bunde-Stanley-2002-PA}.
However, possessing long memory is not sufficient for the precence
of multifractality and one has to have a nonlinear process with
long-memory in order to have multifractality
\cite{Saichev-Sornette-2006-PRE}. In many cases, the null hypothesis
that the reported multifractal nature is stemmed from the large
fluctuations of prices can not be rejected \cite{Lux-2004-IJMPC}.

Here, we propose to investigate the multifractal nature of absolute
returns of stocks based on the multiplier method, again, borrowed
from turbulence
\cite{Chhabra-Sreenivasan-1991-PRA,Chhabra-Sreenivasan-1992-PRL,Jouault-Lipa-Greiner-1999-PRE,Jouault-Greiner-Lipa-2000-PD}.
Our goal is to provide direct evidence of scale invariance in the
distribution of the multipliers. The concept of {\em{multiplier}}
was originally introduced by Novikov to describe the intermittency
and scale self-similarity in turbulent flows
\cite{Novikov-1971-JAMM}. The scale-invariant multiplier
distribution is argued to be more basic than the standard
$f(\alpha)$ curve
\cite{Chhabra-Sreenivasan-1991-PRA,Chhabra-Sreenivasan-1992-PRL}. In
addition, it allows us to extract both positive and negative parts
of the $f(\alpha)$ function with exponentially less computational
time and is more accurate than conventional box-counting methods
\cite{Chhabra-Sreenivasan-1991-PRA,Chhabra-Sreenivasan-1992-PRL}.

\section{Description of the data set}
\label{s1:data}

We adopt a nice high-frequency data set recording the S\&P 500 index
to ensure better statistics in our analysis. The record contains
quoted prices $I(t)$ of the index, covering eighteen years from Jan.
1, 1982 to Dec. 31, 1999. The sampling interval is one minute. As
usual, the nontrading time periods are treated as frozen such that
we count merely the time during trading hours and remove closing
hours, weekends, and holidays from the data. The size of the data
set is about 1.7 million.

The return $r(t)$ over a time scale $\Delta t$ is defined as follows
\begin{equation}
 r(t)=\ln[I(t)]-\ln[I(t-\Delta t)]~,
 \label{Eq:return}
\end{equation}
whose absolute is a measure of the volatility. In this letter, the
time scale is $\Delta t = 1$ min. We can construct an additive
measure in the time interval $[t_1,t_2]$, which is the sum of
absolute returns:
\begin{equation}
 \mu([t_1,t_2]) = \sum_{t = t_1}^{t_2}|r(t)|~.
 \label{Eq:measure}
\end{equation}
The quantity $\mu([t_1,t_2])$ is actually a measure of the
volatility on the time interval $[t_1,t_2]$
\cite{Bollen-Inder-2002-JEF}. The time series $|r(t)|$ is
partitioned into boxes of identical size $s$. Each of these
{\em{mother}} boxes is further divided into $a$ {\em{daughter}}
boxes. The multiplier $m$ is determined by the ratio of the measure
on a daughter box to that on her mother box
\cite{Chhabra-Sreenivasan-1992-PRL}. Therefore, the multiplier is
dependent of $a$ and $s$ and can be denoted as $m_{a,s}$ when
necessary.

\section{Scale invariant distribution}
\label{s1:Prob}

\begin{figure}[htp]
\begin{center}
\includegraphics[width=6.5cm]{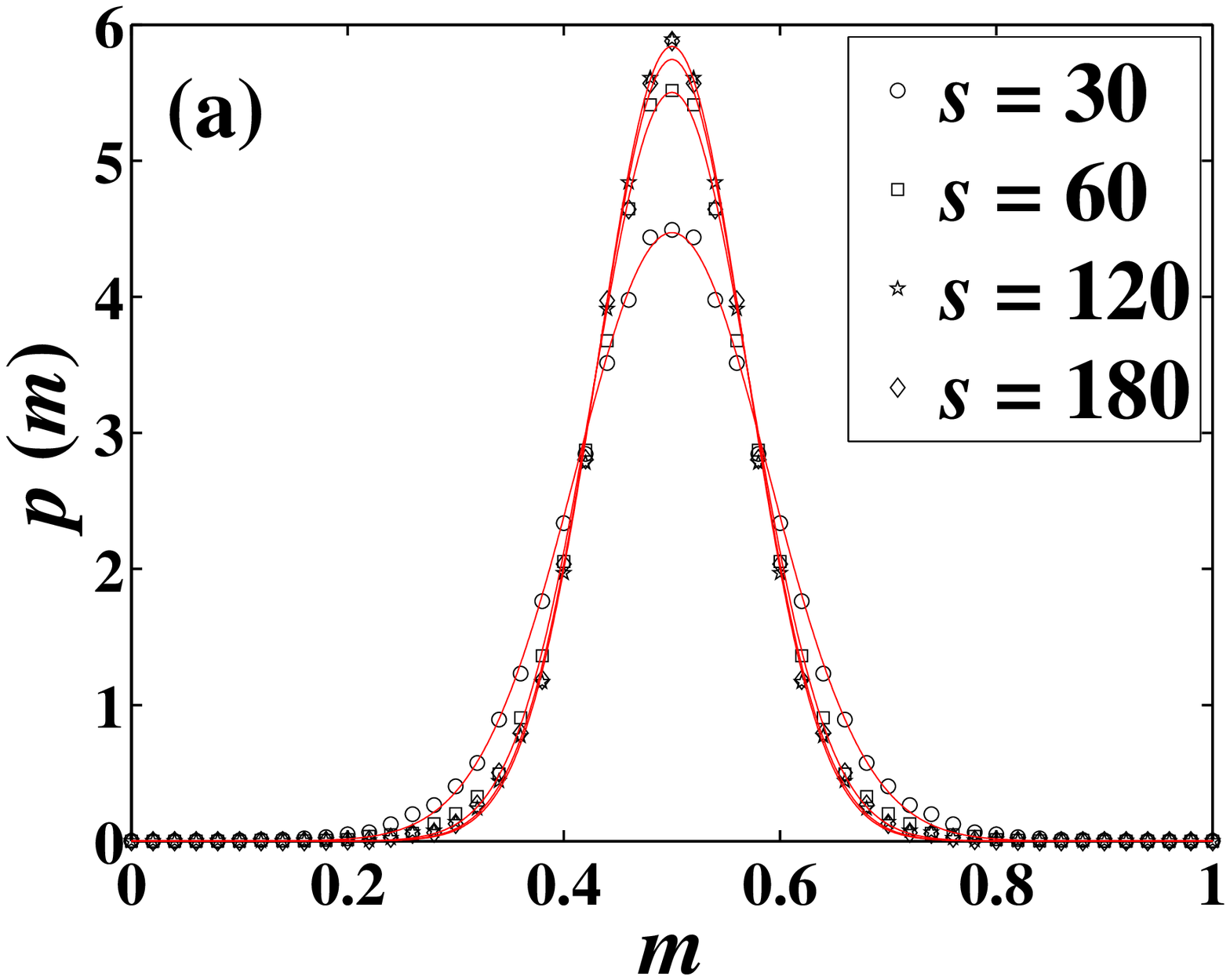}
\includegraphics[width=6.5cm]{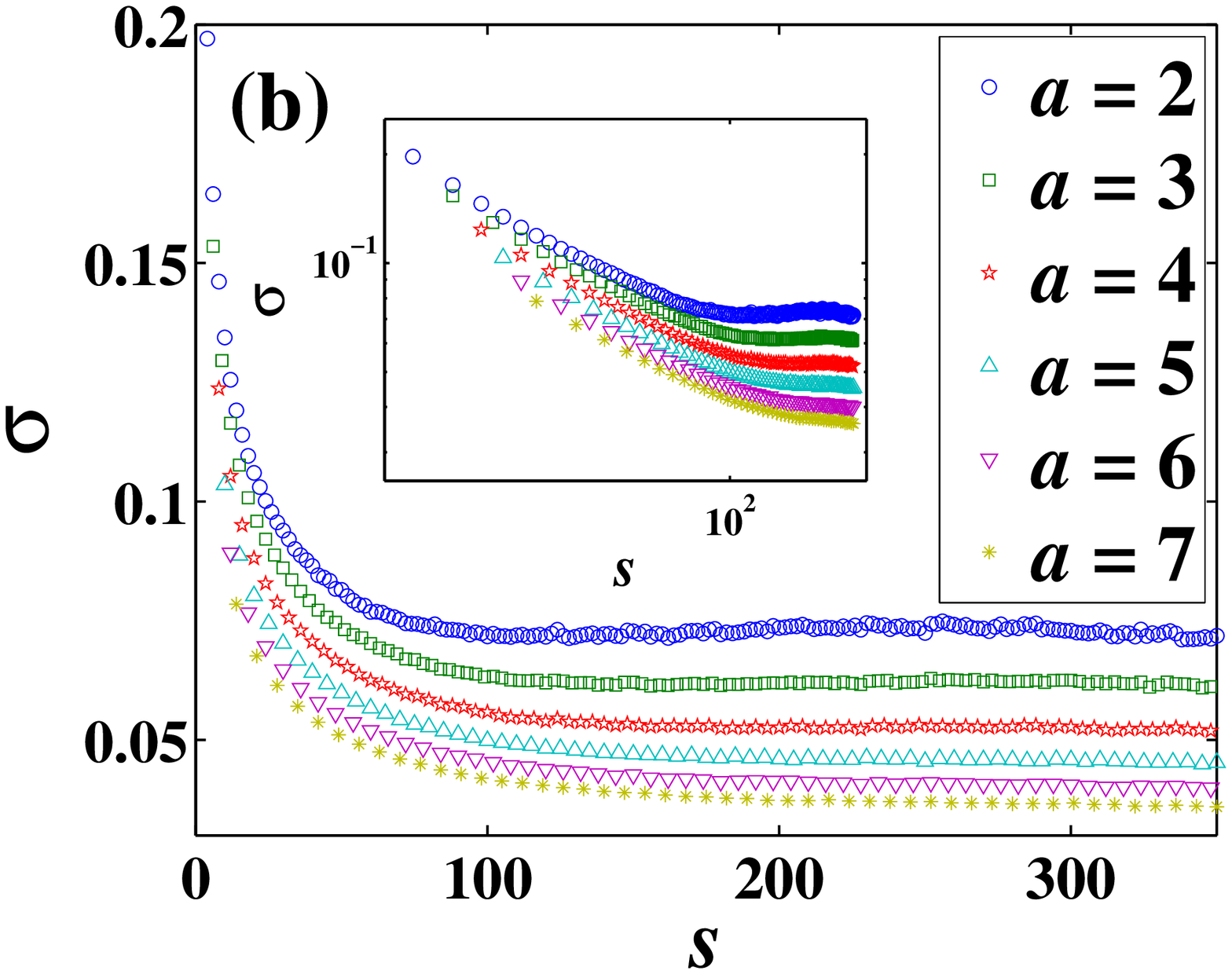}
\caption{(color online). (a) Empirical probability density functions
of the volatility multipliers with the base $a = 2$ for different
box sizes. The solid lines are fitted normal distributions. (b)
Dependence of the sample standard deviation $\sigma$ as a function
of $s$ for different bases. The inset shows the loglog plots of
$\sigma$ against $s$.} \label{Fig:SR}
\end{center}
\end{figure}

Figure~\ref{Fig:SR}(a) presents the probability densities
$p_{a,s}(m)$ of the multiplier $m$ for four different box sizes $s =
30$, $60$, $120$, and $180$ with the same base $a = 2$. All curves
are symmetric with respect to $m = 0.5$ such that
$p_{2,s}(m)=p_{2,s}(1-m)$ by definition and close to Gaussian. The
solid lines are the best fits to normal distributions, whose fitted
standard deviations are $\widehat{\sigma} = 0.089$ for $s = 30$,
$\widehat{\sigma} = 0.073$ for $s = 60$, $\widehat{\sigma} = 0.068$
for $s = 120$, and $\widehat{\sigma} = 0.069$ for $s = 180$,
respectively. The corresponding r.m.s. of the fit residuals are
$0.050$, $0.068$, $0.050$, and $0.068$. Note that the mean $\mu=1/2$
is fixed in the fitting procedure. It is evident that
$\widehat{\sigma}$ decreases in regard to $s$ and tends to a
constant for large $s$. This phenomenon is further manifested by
Fig.~\ref{Fig:SR}(b), which plots the sample standard deviation
$\sigma$ of the multipliers as a function of the box size $s$ for
different base $a$. The inset shows the loglog plots of $\sigma$
against $s$.

One can see that there are two regimes in the $\sigma$ versus $s$
relation: $\sigma$ decays as a power law for small $s$ and saturates
to a constant for large $s$. Roughly speaking, the crossover values
of $s$ are the following: $s_\times \approx 80$ for $a = 2$,
$s_\times \approx 100$ for $a = 3$, $s_\times \approx 120$ for $a =
4$, $s_\times \approx 140$ for $a = 5$, $s_\times \approx 150$ for
$a = 6$, and $s_\times \approx 150$ for $a = 7$, respectively. In
other words, the sample variance $\sigma^2$ in the case of $a=2$ has
the fastest convergence rate to a constant.

\begin{figure}[htp]
\begin{center}
\includegraphics[width=6.5cm]{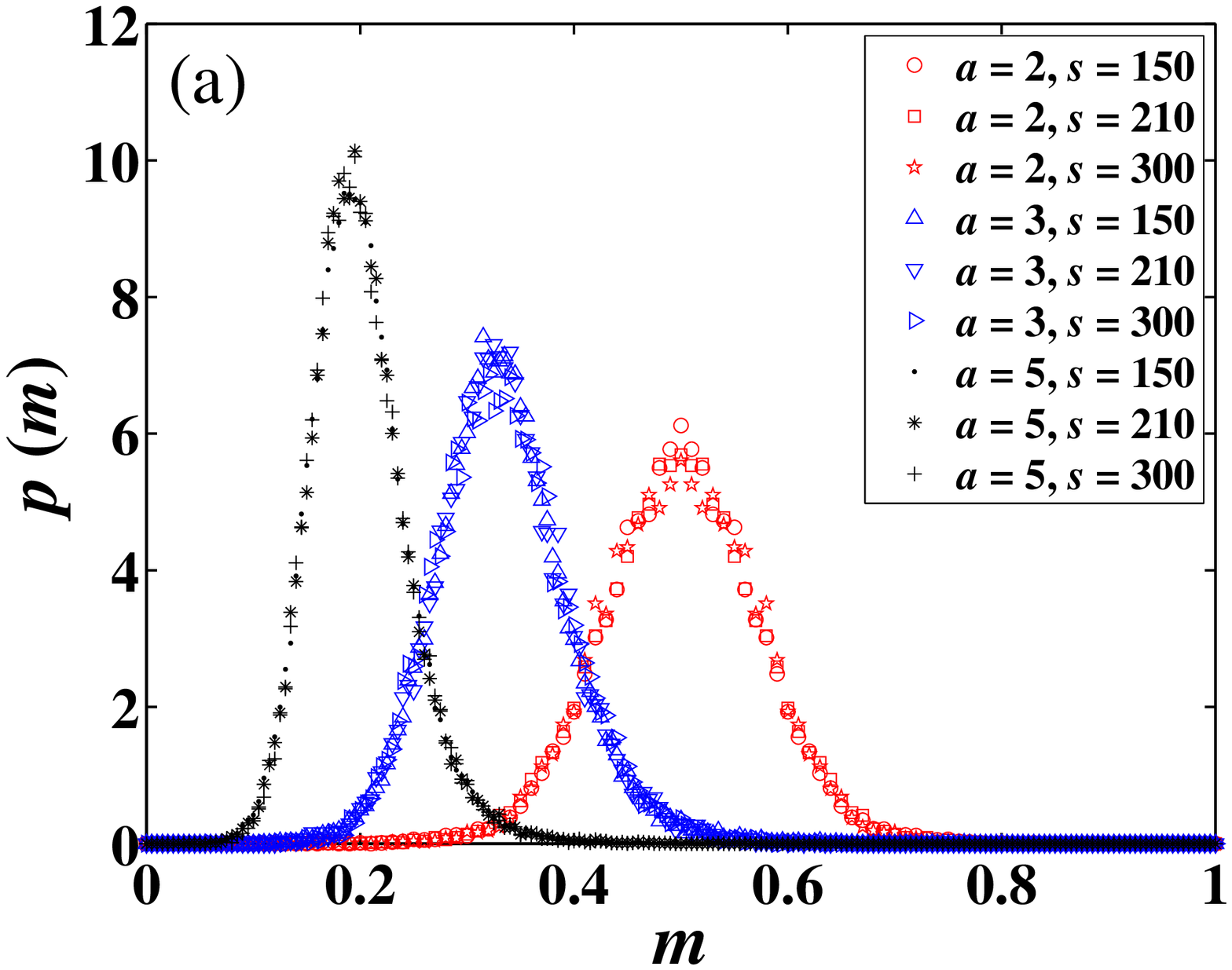}
\includegraphics[width=6.5cm]{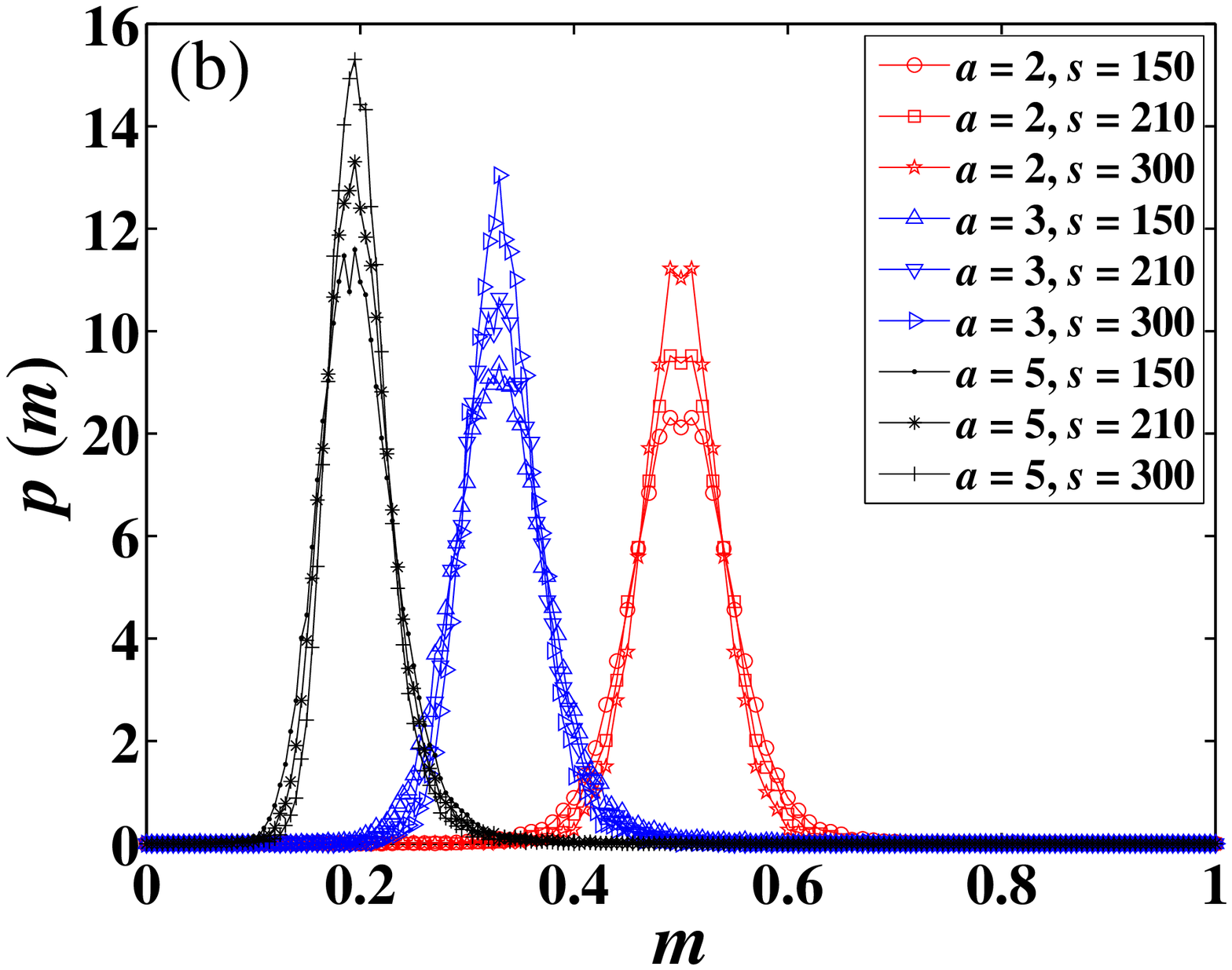}
\caption{(color online). (a) Scale invariant distributions of
volatility multiplier $m_{a,s}$ for different bases $a = 2$, $3$,
and $5$ and different box sizes $m = 150$, $210$, and $300$. (b)
Scale invariant distributions of volatility multiplier $m_{a,s}$ for
the shuffled time series.} \label{Fig:SIPDF}
\end{center}
\end{figure}

Figure~\ref{Fig:SIPDF}(a) shows the empirical probability density
functions $p_{a,s}(m)$ of the multipliers for different bases $a =
2$, $3$, and $5$ and different box sizes $s = 150$, $210$, and
$300$. For each $a$, $p_{a,s}(m)$ remains invariant in respect to
$s$ when $s>s_\times$. In other words, there is a scaling range in
which the volatility multiplier is scale invariant, whose
distribution is independent of $s$. We shuffled the return series
and found that the multiplier distributions are not scale invariant
and the scaling range of $s$ disappears, as illustrated in
Fig.~\ref{Fig:SIPDF}(b). Therefore, $p_{a,s}(m)$ can be reduced to
$p_a(m)$ in the scaling range. The shuffling test shows that long
memory in the volatility plays an essential role in the appearance
of scale invariance.

\begin{figure}[htp]
\begin{center}
\includegraphics[width=6.5cm]{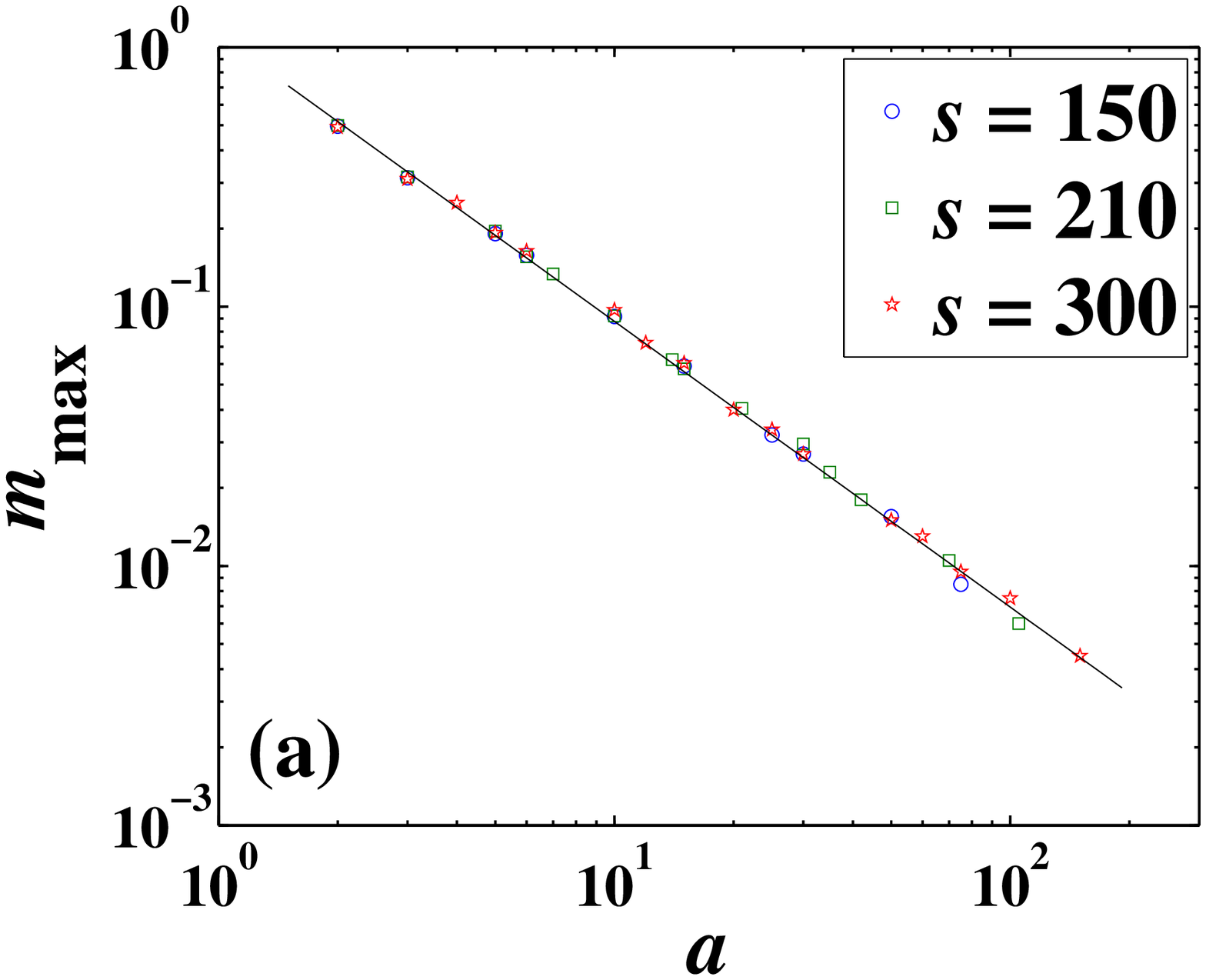}
\includegraphics[width=6.5cm]{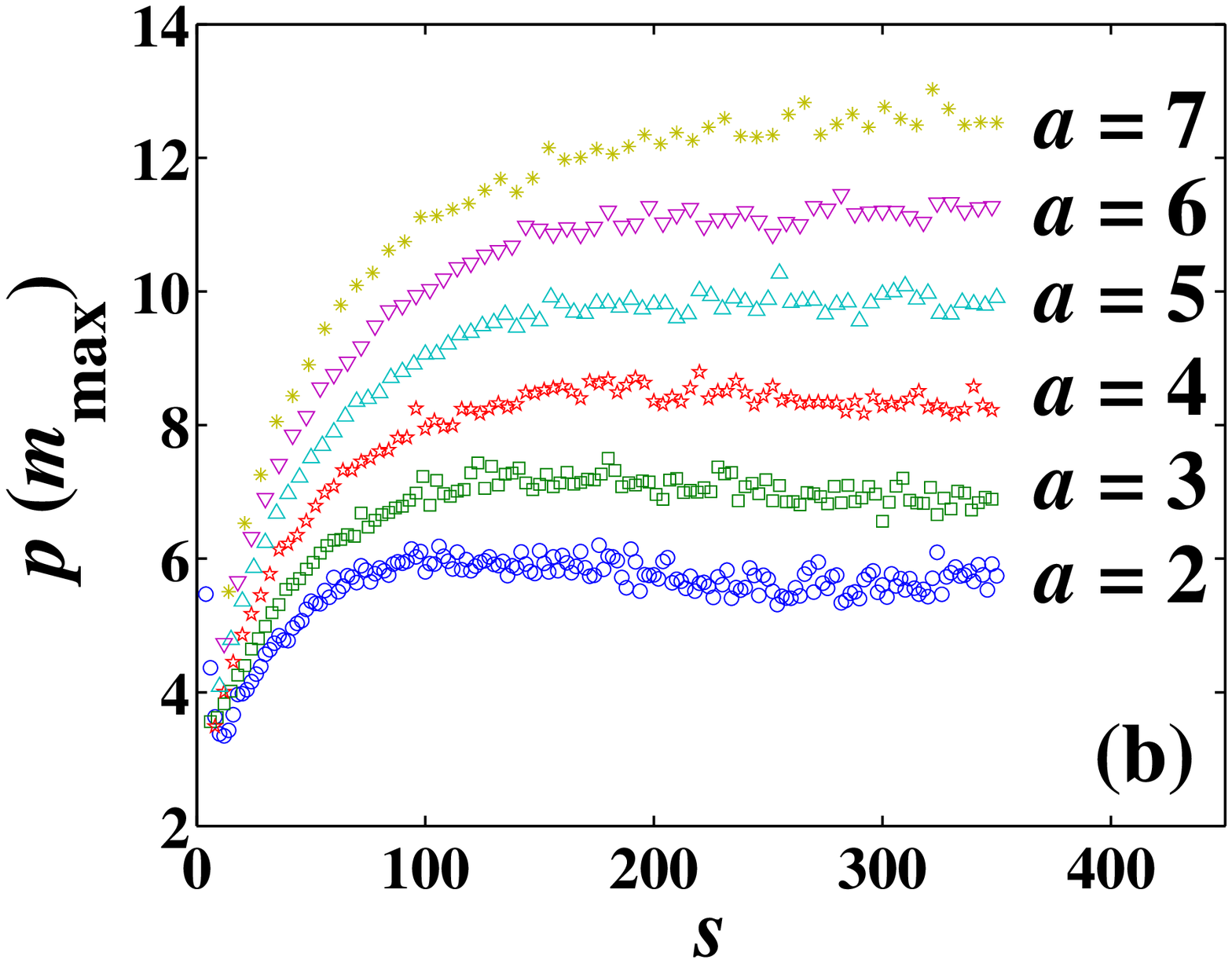}
\caption{(color online). (a) Power law dependence of $m_{\max}$
versus $a$ for $s = 150$, $210$, and $300$. (b) The saturation
behavior of $p(m_{\max})$ in regards to $s$ for different $a$.}
\label{Fig:MP}
\end{center}
\end{figure}

The most probable multipliers are also investigated in this work.
For a given $s$, the most probable multiplier $m_{\max}$ with base
$a$ is estimated such that $p_{a,s}(m_{\max}) = \max[p_{a,s}(m)]$.
Figure~\ref{Fig:MP}(a) presents the loglog plots of $m_{\max}$
versus $a$ for different values of $s$. One can observe that the
data points for different $s$ collapse on a single line, showing a
power-law dependence
\begin{equation}
 m_{\max} \approx a^{-\beta}~,
 \label{Eq:PL}
\end{equation}
where $\beta = 1.10 \pm 0.02$ for $s = 150$, $\beta = 1.10 \pm 0.01$
for $s = 210$, and $\beta = 1.09 \pm 0.01$ for $s =300$,
respectively. Intuitively, since a mother box is divided into $a$
daughter boxes, the sum of the $a$ multipliers is one and the
multipliers is expected to aggregate around $1/a$. However, it is
noteworthy that this power-law dependence is nontrivial, which does
not hold in turbulence
\cite{Chhabra-Sreenivasan-1992-PRL,Molenaar-Herweijer-vandeWater-1995-PRE,Kluiving-Pasmanter-1996-PA}.
In Fig.~\ref{Fig:MP}(b) is shown $p_{a,s}(m_{\max})$ as a function
of $s$ for different $a$. It is evident that $p(m_{\max})$ increases
with $s$ and then approaches a plateau when $s>s_\times$. Figure
\ref{Fig:MP} further verifies the scale-invariant nature of the
volatility multiplier.

\section{Multifractal analysis}
\label{s1:MF}

It is shown that, for any two bases $a$ and $b$, the density
functions are related through Mellin transform in the following form
\cite{Chhabra-Sreenivasan-1992-PRL}
\begin{equation}
 [\mathbb{M} \{ p_a(m_a) \}]^{1 / \ln a} = [\mathbb{M} \{ p_b(m_b) \}]^{1 / \ln
 b}~,
 \label{Eq:Mellin}
\end{equation}
where $\mathbb{M}$ stands for Mellin transform. Equivalently, we
have
\begin{equation}
 \frac{\ln \int_0^1 m_a^q p_a(m_a) d m_a}{\ln a} = \frac{\ln \int_0^1 m_b^q p_b(m_b) d m_b}{\ln
 b}~.
 \label{Eq:base}
\end{equation}
The scaling exponent $\tau(q)$ of the moment of $m_a$ can be
obtained as follows
\cite{Chhabra-Sreenivasan-1991-PRA,Chhabra-Sreenivasan-1992-PRL}
\begin{equation}
 \tau(q) = - D_0 - \frac{\ln\langle m_a^q \rangle}{\ln a}~,
 \label{Eq:tau}
\end{equation}
where $D_0$ is the fractal dimension of the support of the measure.
In the current case, we have $D_0 = 1$. Note that the use of the
Mellin transform may indeed appear natural in the framework of
scaling and power-like functions, for instance, in the analysis of
Weierstrass-type functions \cite{Gluzman-Sornette-2002-PRE}.

The local singularity exponent $\alpha$ and its spectrum $f(\alpha)$
are related to $\tau(q)$ through Legendre transforms: $\alpha(q) =
\tau'(q)$ and $f(\alpha) = q \alpha(q) - \tau(q)$. It follows that
\cite{Chhabra-Sreenivasan-1991-PRA,Chhabra-Sreenivasan-1992-PRL,Zhou-Yu-2001-PA,Zhou-Liu-Yu-2001-Fractals,Zhou-Yu-2001-PRE}
\begin{equation}
 \alpha(q) = -\frac{\langle m_a^q \ln m_a \rangle}{\langle m_a^q \rangle \ln a}
 \label{Eq:alpha}
\end{equation}
and
\begin{equation}
 f(\alpha) =  \frac{\langle m_a^q \rangle \ln \langle m_a^q \rangle
 - \langle m_a^q \ln m_a^q \rangle}{\langle m_a^q \rangle \ln
 a}~.
 \label{Eq:falpha}
\end{equation}
Equations (\ref{Eq:base}-\ref{Eq:falpha}) predict that each of these
characteristic multifractal arguments for fixed $q$ converges to
constant in the scaling range and are independent of the base $a$ as
well.

\begin{figure}[htb]
\begin{center}
\includegraphics[width=8cm]{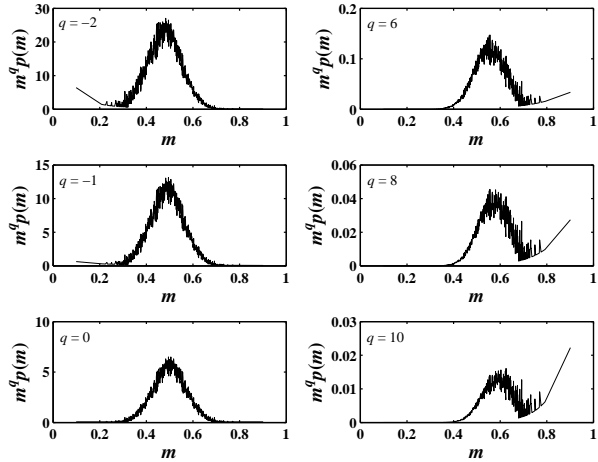}
\caption{Numerical integrand $m^q p(m)$ as a function of $m$ for
different values of $q$ with $s = 210$ and $a = 2$.} \label{Fig:C}
\end{center}
\end{figure}

In order to test this prediction, we have to calculate first the
scaling function $\tau(q)$ which requires that the integrand
$m_a^qp_a(m_a)$ converges for a given $q$
\cite{Lvov-Podivilov-Pomyalove-Procaccia-Vandembroucq-1998-PRE,Zhou-Sornette-Yuan-2006-PD}.
We have investigated $m_a^qp_a(m_a)$ for different values of $q$,
$s$, and $a$. A typical dependence of $m_a^qp_a(m_a)$ as a function
of $m_a$ is shown in Fig.~\ref{Fig:C} for different values of $q$
with fixed box size $s = 210$ and base $a=2$. The integrand diverges
for large $m_a$ when $q$ is larger than 6. We shall nevertheless
investigate scaling functions for $q\leqslant8$ for comparison.
Moreover, Fig.~\ref{Fig:C} indicates that the integrand diverges
when $q\leqslant-1$ and the associated negative moments do not
exist. This is a direct consequence of the fact that $p_a(0)\neq0$.
Indeed, there are time moments when the local returns are zero so
that the probability density at $m_a=0$ is apart from zero,
{\em{i.e.}}, $p_a(0)\neq0$. Approximately, for a small number
$\delta$, $p(m)=p(0)$ is a constant for $m<\delta$. Posing
$\int_{\delta}^1m^qp(m)dm = C$, we have
$\int_0^1m^qp(m)dm=\int_0^{\delta}m^qp(m)dm+ C
=p(0)\frac{1}{q+1}m^{q+1}\left|\right._0^{\delta}+C$. Therefore, we
have $q>-1$, which is quite analogous to the situation in turbulence
\cite{Pearson-vandeWater-2005-PRE}.

\begin{figure}[htb]
\begin{center}
\includegraphics[width=8cm]{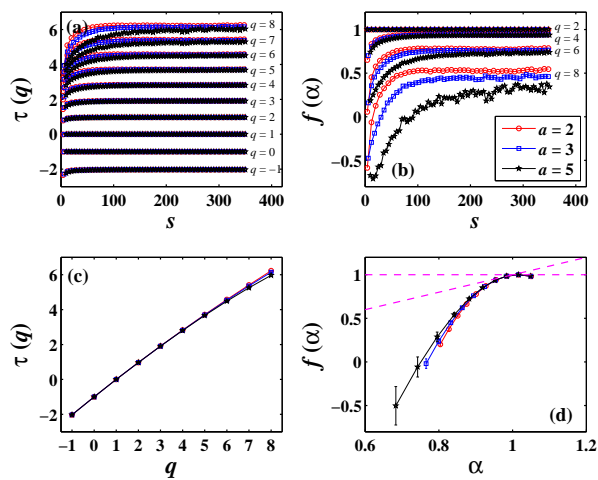}
\caption{(color online). Multifractal analysis of absolute returns.
(a) Dependence of $\tau (q)$ with respect to $s$ for different
values of $q$ and different bases $a = 2$, $3$, and $5$. (b)
Dependence of $f(\alpha)$ with respect to $s$ for different values
of $q$ and different bases $a = 2$, $3$, and $5$. (c) Scaling
function $\tau (q)$ obtained from Eq.~(\ref{Eq:tau}) for different
bases $a = 2$, $3$, and $5$. (d) Singularity spectrum $f(\alpha)$
obtained from Eq.~(\ref{Eq:alpha}) and Eq.~(\ref{Eq:falpha}) for
different bases $a = 2$, $3$, and $5$.}\label{Fig:MF}
\end{center}
\end{figure}

We now present in Fig.~\ref{Fig:MF} the results of the multifractal
analysis for $q$ varying from $-1$ to $8$. In Fig.~\ref{Fig:MF}(a)
is shown the dependence of the scaling function $\tau(q)$ upon the
box size $s$ for different values of $q$ and $a$. It is evident
that, for large $s$, $\tau(q)$ is independent of $s$ for every $q$
under investigation. The $\tau(q)$ function reaches constant faster
for small $a$. These results are in excellent agreement with the
theoretical predictions. Figure~\ref{Fig:MF}(b) shows the dependence
of the singularity spectrum $f(\alpha)$ on the box size $s$ for
different values of $q$ and $a$. Again, we witness a range of scale
invariance in which $f(\alpha)$ is independent of $s$. What needs to
be emphasized is that, for $q=8$, the three $f(\alpha)$ curves with
different $a$ do not converge due to bad statistics as shown by the
left-middle panel of Fig.~\ref{Fig:C}. We plot the three scaling
functions $\tau(q)$ for $a=2$, $3$, and $5$ with respect to $q$ in
Fig.~\ref{Fig:MF}(c). The error bars are estimated as the standard
deviation over different $s$. Except for large $q$, the three
$\tau(q)$ curves collapse on a single nonlinear curve. In addition,
Fig.~\ref{Fig:MF}(d) shows the three singularity spectra $f(\alpha)$
in respect to the local singularity exponent $\alpha$ for the three
bases. Again, the three curves collapse remarkably on a single curve
when $q$ is not too large. Both Fig.~\ref{Fig:MF}(c) and
Fig.~\ref{Fig:MF}(d) strongly indicate that the volatility
multiplier possesses multifractal nature.

An important feature of multifractals is the possible existence of
negative dimensions in the multifractal spectrum, that is,
$f(\alpha) < 0$ for large or small $\alpha$
\cite{Mandelbrot-1990a-PA}. Negative dimensions are more common if
the multiplier distribution is continuous
\cite{Chhabra-Sreenivasan-1992-PRL,Zhou-Yu-2001-PA,Zhou-Liu-Yu-2001-Fractals,Zhou-Yu-2001-PRE}.
Figure \ref{Fig:MF}(d) also shows that there are negative dimensions
for large $q$ especially when $a$ is large. However, we should be
cautious that the part of $f(\alpha)<0$ might be an artifact of bad
statistics for large $q$ and $a$, as shown by the right panel of
Fig.~\ref{Fig:C} and Fig.~\ref{Fig:MF}(b) as well. Since the
multifractal functions are more reliable statistically for $a=2$, we
argue that there is no negative dimension for $q\leqslant8$. More
data are required to investigate higher order moments and the issue
of the existence of negative dimensions is still open.

In the development of Econophysics, the literature has witnessed
increasing analogues between turbulence and finance. Multiplier
analysis is a well-established method in the description of
conservative quantities in turbulence
\cite{Novikov-1971-JAMM,Chhabra-Sreenivasan-1991-PRA,Chhabra-Sreenivasan-1992-PRL,Jouault-Lipa-Greiner-1999-PRE,Jouault-Greiner-Lipa-2000-PD}.
Novikov predicted that the multiplier distribution $p(m)$ is
independent of the scale $s$ as long as $s$ is well inside the
inertial subrange \cite{Novikov-1971-JAMM}. Our finding provides
further evidence of the analogue between turbulence and finance,
except that the existence of negative dimensions that was reported
in turbulence is not confirmed in our financial data. Moreover, the
energy dissipation multiplier with $a=2$ follows approximately a
triangular distribution, which is much flatter than the normal
distribution of volatility multiplier in this work.

\section{Conclusion}

In summary, we have employed the multiplier method to investigate
the volatility of high-frequency data of the S\&P 500 index. The
distribution of volatility multiplier is found to be independent of
the time scale $s$ for different $a$ when $s$ is larger than some
crossover scale $s_\times$. We unraveled that the volatility
multipliers are scale invariant and have multifractal nature, which
is independent of the construction of the multipliers (characterized
by $s$ and $a$) in the scaling range.

\bigskip
{\textbf{Acknowledgments:}}

This work was partially supported by the National Natural Science
Foundation of China (Grant No. 70501011) and the Fok Ying Tong
Education Foundation (Grant No. 101086).

\bibliographystyle{elsart-num}
\bibliography{E:/papers/Bibliography}

\end{document}